%% file: author.tex
\documentclass[graybox]{svmult}

\usepackage{mathptmx}       % selects Times Roman as basic font
\usepackage{helvet}         % selects Helvetica as sans-serif font
\usepackage{courier}        % selects Courier as typewriter font
\usepackage{type1cm}        % activate if the above 3 fonts are
\usepackage{makeidx}         % allows index generation
\usepackage{graphicx}        % standard LaTeX graphics tool
\usepackage{multicol}        % used for the two-column index
\usepackage[bottom]{footmisc}% places footnotes at page bottom

\makeindex             % used for the subject index
\begin{document}

\title*{New Physics solutions for $b\rightarrow c\,\tau\,\bar{\nu}$ anomalies after Moriond 2019}
\titlerunning{New Physics solutions for $b\rightarrow c\,\tau\,\bar{\nu}$ anomalies: Post-Moriond 2019} %for an abbreviated version of
\author{Suman Kumbhakar, Ashutosh Kumar Alok, Dinesh Kumar and S Uma Sankar}
\institute{Suman Kumbhakar \at Indian Institute of Technology Bombay, Mumbai 400076, India, \email{suman@phy.iitb.ac.in}
\and Ashutosh Kumar Alok \at Indian Institute of Technology Jodhpur, Jodhpur 342011, India, \email{akalok@iitj.ac.in}
\and Dinesh Kumar \at University of Rajasthan, Jaipur 302004, India,\\ National Centre for Nuclear Research, Warsaw, Poland, \email{dinesh@uniraj.ac.in}
\and S Uma Sankar \at Indian Institute of Technology Bombay, Mumbai 400076, India, \email{uma@phy.iitb.ac.in}}
\maketitle

%\abstract*{}

\abstract{At Moriond 2019, Belle collaboration has announced new measurements on the flavour ratios $R_D - R_{D^*}$ which are consistent with their Standard Model predictions within $1.2\sigma$. After inclusion of these measurements, the global tension in $R_D - R_{D^*}$ has reduced from $4.1\sigma$ to $3.1\sigma$ which is still significant. The measurements of these ratios indicate towards the violation of lepton flavor universality in $b\rightarrow c\,l\,\bar{\nu}$ decay. Assuming new physics in $b\rightarrow c\,\tau\,\bar{\nu}$ transition, we have done a global fit to all available data in this sector to identify the allowed new physics solutions. We find that there are seven allowed new physics solutions which can account for all measurements in $b\rightarrow c\,\tau\,\bar{\nu}$ transition. We show that a simultaneous measurement of the $\tau$ polarization fraction and forward-backward asymmetry in $B\rightarrow D\,\tau\,\bar{\nu}$, the zero crossing point of forward backward asymmetry in $B\rightarrow D^*\tau\bar{\nu}$ and the branching ratio of $B_c\rightarrow \tau\,\bar{\nu}$ decay can distinguish these seven new physics solutions if they can be measured with a required precision.}
%%%%%%%%%%%%%%%%%%%%%%%%%%%%%%%%%%%%
\section{Introduction}
\label{sec:1}
%%%%%%%%%%%%%%%%%%%%%%%%%%%%%%%%%
In recent years, several measurements in B meson decays, reported by LHCb and B-factories, show significant tension with their Standard Model (SM) predictions. One class of such measurements is governed by $b\rightarrow c\,l\,\bar{\nu}$ transitions. This transition occurs at tree level within the SM. The BaBar, Belle and LHCb experiments made a series of measurements of the flavour ratios
\begin{equation}
R_{D} = \frac{\mathcal{B}(B\rightarrow D\,\tau\,\bar{\nu})} {\mathcal{B}(B\rightarrow D\, l \, \bar{\nu})},\hspace{0.1cm}
R_{D^*} = \frac{\mathcal{B}(B\rightarrow D^{*}\,\tau\,\bar{\nu})} {\mathcal{B}(B\rightarrow D^{*}\, l \, \bar{\nu})}, \hspace{0.1cm} l= e \,{\rm or}\, \mu .
\label{rdrds}
\end{equation}
The discrepancy between the world average of 2018 and the SM prediction was at a level of $\sim 4\sigma$~\cite{Amhis:2016xyh}. Very recently, Belle collaborations made new measurements of these ratios which are consistent with their SM predictions within $\sim 1.2\sigma$~\cite{Abdesselam:2019dgh}. These results were announced at Moriond 2019. After inclusion of these new measurements, the present world averges of these ratios are about $\sim 3.1\sigma$ away from the SM predictions~\cite{avg19}. These ratios indicate towards lepton flavour universality violation.

In 2017, LHCb measured another flavour ratio $R_{J/\psi} = \mathcal{B}(B_c\rightarrow J/\psi\,\tau\,\bar{\nu})/\mathcal{B}(B_c\rightarrow J/\psi\,\mu\,\bar{\nu})$ and found it to be $\sim 1.7\sigma$ higher than the SM prediction~\cite{Aaij:2017tyk}. In addition to these flavour ratios, Belle collaboration has measured two angular observables in $B\rightarrow D^*\,\tau\,\bar{\nu}$ decay $-$ the $\tau$ polarization fraction $P^{D^*}_{\tau}$~\cite{Hirose:2016wfn} and the longitudinal polarization fraction of $D^*$ meson $f^{D^*}_{L}$~\cite{Adamczyk:2019wyt}. The measured value of $P^{D^*}_{\tau}$ is consistent with the SM prediction because it has a very large statistical error. However, the measured value of $f_L^{D^*}$ is about $\sim 1.6\sigma$ higher than the SM prediction.

The discrepancy in $R_D$ and $R_{D^*}$ could be an indication of presence of new physics (NP) in the $b\rightarrow c\tau\bar{\nu}$ transition. The possibility of NP in $b\rightarrow c\{e/\mu\}\bar{\nu}$ is excluded by other data~\cite{Alok:2017qsi}. All possible NP four-Fermi operators for $b\rightarrow c\tau\bar{\nu}$ transition are listed in ref.~\cite{Freytsis:2015qca}. In ref~\cite{Alok:2017qsi}, a fit was performed to all the $b\rightarrow c\tau\bar{\nu}$ data available up to summer 2018. It was found that there are six allowed NP solutions. Among those six solutions, four solutions are distinct each with different Lorentz structure. In ref~\cite{Alok:2018uft}, we have shown that a unique discrimination between the allowed NP solutions can be possible by a simultaneous measurements of four angular observables, $P_{\tau}^{D^*}$ ($\tau$ polarization fraction), $f^{D^*}_{L}$ (longitudinal $D^*$ polarization fraction), $A^{D^*}_{FB}$ (the forward-backward asymmetry), $A^{D^*}_{LT}$ (longitudinal-transverse asymmetry) in the $B\rightarrow D^*\tau\bar{\nu}$ decay~\cite{Kumbhakar:2019gry}.

In this work, we study the impact of new Belle measurements of $R_D$-$R_{D^*}$ and $f^{D^*}_{L}$ on the NP solutions for $b\rightarrow c\tau\bar{\nu}$ anomalies. We redo the global fit by taking all present measurements in this sector and find out the presently allowed NP solutions~\cite{Alok:2019uqc,Kumbhakar:2019avh}. We also discuss methods to discriminate between the allowed NP solutions by means of angular observables in $B\rightarrow (D,D^*)\tau\bar{\nu}$ decays and the branching ratio of $B_c\rightarrow \tau\bar{\nu}$ decay.
%%%%%%%%%%%%%%%%%%%%%%%%%%%%%%%%%%%
\section{New Physics solutions after Moriond 2019}
\label{sec:2}
%%%%%%%%%%%%%%%%%%%%%%%%%%%%%%%%%%%%%%%%
The most general effective Hamiltonian for $b\rightarrow c\tau\bar{\nu}$ transition can be written as
\begin{equation}
H_{eff}= \frac{4 G_F}{\sqrt{2}} V_{cb}\left[O_{V_L} + \frac{\sqrt{2}}{4 G_F V_{cb}} \frac{1}{\Lambda^2} \left\lbrace \sum_i \left(C_i O_i +
 C^{'}_i O^{'}_i + C^{''}_i O^{''}_i \right) \right\rbrace \right],
\label{effH}
\end{equation}
where $G_F$ is the Fermi coupling constant, $V_{cb}$ is the Cabibbo-Kobayashi-Maskawa (CKM) 
matrix element and the NP scale $\Lambda$ is assumed to be 1 TeV. We also assume that neutrino is always left chiral. The effective Hamiltonian for the SM contains only the $O_{V_L}$ operator. The explicit forms of the four-fermion operators $O_i$, $O^{'}_i$ and $O^{''}_i$ are given in ref~\cite{Freytsis:2015qca}.
The NP effects are encoded in the NP Wilson coefficients (WCs) $C_i, C^{'}_i$ and $C^{''}_i$. 
Each primed and double primed operator can be expressed as a linear combination of unprimed operators through Fierz transformation.

First we fit the NP predictions of $R_D$, $R_{D^*}$, $R_{J/\psi}$, $P_{\tau}^{D^*}$ and $f_L^{D^*}$ to the current measured values. The corresponding $\chi^2$ is defined as
\begin{equation}
\chi^2(C_i)= \sum \left(O^{\rm th}(C_i)-O^{\rm exp}\right)^{T} \mathcal{C}^{-1} \left(O^{\rm th}(C_i)-O^{\rm exp}\right).
\label{chi2}
\end{equation}
Here $\mathcal{C}$ is the covariance matrix which includes both theory and experimental correlations. The fit is done by using the CERN minimization code {\tt MINUIT}~\cite{James:1994vla}. We perform three types of fits: (a) taking only one NP operator at a time, (b) taking two similar NP operators at a time, (c) taking two dissimilar NP operators at a time. The branching ratio of $B_c\rightarrow \tau\bar{\nu}$ puts a stringent constraint on the scalar/pseudo-scalar NP WCs. In particular, LEP data imposes an upper bound on this quantity which is $\mathcal{B}(B_c\rightarrow \tau\bar{\nu})<10\%$~\cite{Akeroyd:2017mhr}. Further we include the renormalization group (RG) effects in the evolution of the WCs from the scale $\Lambda= 1$ TeV to
the scale $m_b$~\cite{Gonzalez-Alonso:2017iyc}. 
%%%%%%%%%%%%%%%%%%%%%%%%%%%
 \begin{table}[htbp]
 \centering
 \tabcolsep 7pt
 \begin{tabular}{|c|c|c|}
\hline\hline
NP type & Best fit value(s)& $\chi^2_{\rm min}$  \\
\hline
SM  & $C_{i}=0$ & $21.80$  \\
\hline
$C_{V_L}$  &  $0.10 \pm 0.02$& $4.5$ \\

\hline
$(C''_{S_L},\, C''_{S_R})$  & $(0.05, 0.24)$ &$4.4$   \\

\hline
$(C_{S_L}, C_T)$ & $(0.06, -0.06)$ & 5.0 \\
\hline
$(C_{S_R}, C_T)$ & $(0.07, -0.05)$ & 4.6 \\
\hline
$(C''_{V_R}, C''_{T})$ & $(0.21, 0.11)$ & 4.2 \\
\hline\hline	

$C_T$ & $-0.07\pm 0.02$ & $7.1$ \\

\hline
$(C'_{V_R},\, C'_{S_L})$  & $(0.38, 0.63)$  &$6.0$  \\
\hline
$(C''_{V_R},\, C''_{S_L})$  & $(0.11, -0.58)$ &$6.2$  \\
%$C''_{S_L}$ & $-0.22\pm 0.02$ & $9.2$ & 0.3 \\

\hline\hline
\end{tabular}
\caption{Fit values of the coefficients of new physics operators at $\Lambda = 1$ TeV by making use of data of $R_D$, $R_{D^*}$, $R_{J/\psi}$, $P_{\tau}^{D^*}$ and $f_L^{D^*}$. In this fit, we use the HFLAV summer 2019 averages of $R_D$-$R_{D^*}$. All new physics solutions satisfy $\chi^2_{\rm min}\leq 7$ as well as $\mathcal{B}(B_c\rightarrow \tau\,\bar{\nu})< 10\%$.}
\label{tab1}
 \end{table}
%%%%%%%%%%%%%%%%%%%%%%%%%%%%%%%%%%%%%%

The  $B\rightarrow D^{(*)}\, l\,  \bar{\nu}$ decay distributions depend upon hadronic form-factors. The form factors for $B\rightarrow D$ decay are well known in lattice QCD \cite{Aoki:2016frl} and we use them in our analyses. For $B\rightarrow D^*$ decay, the HQET parameters are extracted using data from Belle and BaBar experiments along with lattice inputs. In this work, the numerical values of these parameters are taken from refs. \cite{Bailey:2014tva} and  \cite{Amhis:2016xyh}.

The best fit solutions are listed in table~\ref{tab1} which satisfy the constraints  $\chi^2_{\rm min}\leq 5$ as well as $\mathcal{B}(B_c\rightarrow \tau\bar{\nu})< 10\%$. Comparing with the previously allowed solutions (table 4 in ref.~\cite{Alok:2017qsi}), we note that only the $\mathcal{O}_{V_L}$ solution survives among the single operator solutions. However, its coefficient is reduced by a third because of the reduction in discrepancy. Among the two similar operator solutions, only the $(\mathcal{O}''_{S_L},\, \mathcal{O}''_{S_R})$ persists in principle, with the WCs $(C''_{S_L},\, C''_{S_R}) = (0.05, 0.24)$. The value of $C''_{S_L}$ is quite small, $C''_{S_R} \approx 2C_{V_L}$ and the Fierz transform of $\mathcal{O}''_{S_R}$ is $\mathcal{O}_{V_L}/2$. Therefore, this solution is effectively equivalent to the $\mathcal{O}_{V_L}$ solution. In ref.~\cite{Alok:2016qyh}, we have shown that $f^{D^*}_L$ can strongly discriminate against the tensor and scalar NP solutions. The tensor solution $C_T = 0.516$, which was allowed before $f^{D^*}_L$ measurement~\cite{Alok:2017qsi}, is now completely ruled out at the level of $\sim 5\sigma$. The $O''_{S_L}$ solution is now ruled out in view of goodness of fit. 
Table~\ref{tab1} also lists three other solutions with $5\leq \chi^2_{\rm min}\leq 7$. We consider these solutions because the minimum $\chi^2$ is just a little larger than $5$. Hence, they are only mildly disfavoured compare to the five solutions listed above them. One important point to note is that the prediction of $R_D$ (see Table III in ref.~\cite{Alok:2019uqc}) for the tensor NP solution $C_T = -0.07$ is $1.5\sigma$ below the present world average. Hence there are seven NP solutions which can account for the present data in $b\rightarrow c\,\tau\,\bar{\nu}$ transition. After Moriond 2019, several groups have done similar analysis which can be found in refs.~\cite{Hu:2018veh,Murgui:2019czp,Shi:2019gxi,Blanke:2019qrx}.

%%%%%%%%%%%%%%%%%%%%%%%%%%%%
\section{Methods to distinguish new physics solutions}
\label{sec:3}
%%%%%%%%%%%%%%%%%%%%%%%%%%
In order to distinguish between these seven NP solutions, we consider the angular observables in $B\rightarrow (D,D^*) \,\tau \,\bar{\nu}$ decays. We consider the following four observables:
(i) The $\tau$ polarization $P_{\tau}^D$ in $B\rightarrow D \,\tau \,\bar{\nu}$, (ii) The forward-backward asymmetry $A^{D}_{FB}$ in $B\rightarrow D \,\tau\, \bar{\nu}$, (iii) The zero crossing point (ZCP) of $A^{D^*}_{FB}(q^2)$ in $B\rightarrow D^* \,\tau\, \bar{\nu}$ and (iii) The branching ratio of $B_c\rightarrow \tau\,\bar{\nu}$. The predictions of each of these quantities for each of the seven solutions are listed in table~\ref{tab2}.

%%%%%%%%%%%%%%%%
\begin{table}
\centering
\tabcolsep 7pt
\resizebox{\textwidth}{!}{ 
\begin{tabular}{|c|c|c|c|c|c|}
\hline\hline
NP type  & $P_{\tau}^D$ & $A^{D}_{FB}$& $A^{D^*}_{FB}$ &  $\mathcal{B}(B_c\rightarrow \tau\bar{\nu} )\, \%$ & ZCP of $A^{D^*}_{FB}(q^2)$ GeV$^2$ \\
\hline
SM & $0.324\pm 0.001$& $0.360\pm 0.001$ & $-0.012\pm 0.007$  & 2.2 & 5.8\\
\hline
$C_{V_L}$  & $0.324\pm 0.002$ & $0.360\pm 0.002$ &$-0.013\pm 0.007$  &2.5 & 5.8\\
\hline
$(C_{S_L}, C_T)$ & $0.442\pm 0.002$ & $0.331\pm 0.003$ & $-0.069\pm 0.009$ & 0.8 & 7.0\\
\hline
$(C_{S_R}, C_T)$ & $0.450\pm 0.003$ & $0.331\pm 0.002$ & $-0.045\pm 0.007$  &4.0 & 6.4 \\
\hline
$(C''_{V_R}, C''_{T})$ & $0.448\pm 0.002$ & $-0.244\pm 0.003$ & $-0.025\pm 0.008$  & 11.0 & 6.0\\
\hline
\hline
$C_T$ & $0.366\pm 0.003$ & $0.341\pm 0.002$ & $-0.067\pm 0.011$ &  1.9 & 7.0\\

\hline
$(C'_{V_R},\, C'_{S_L})$  &$0.431\pm 0.002$ & $-0.216\pm 0.004$ & $-0.120\pm 0.009$  & 5.7 & 8.6 \\

\hline
$(C''_{V_R},\, C''_{S_L})$  & $0.447\pm 0.003$ & $0.331\pm 0.003$ & $-0.123\pm 0.010$ & 8.4 & 8.6 \\
\hline
\hline
\end{tabular}
}
\caption{The predictions of $P^D_{\tau}$, $A_{FB}^D$, $A^{D^*}_{FB}$, $\mathcal{B}(B_c\rightarrow \tau\,\bar{\nu})$ and the zero crossing point (ZCP) of $A_{FB}^{D^*}(q^2)$ for each of the allowed NP solutions.}
\label{tab2}
\end{table}
%%%%%%%%%%%%%%%%%%%%%%%
%%%%%%%%%%%%%%%%%%%%%%%
\begin{figure}[htbp]
%[htbp] 
\centering
\includegraphics[scale=0.5]{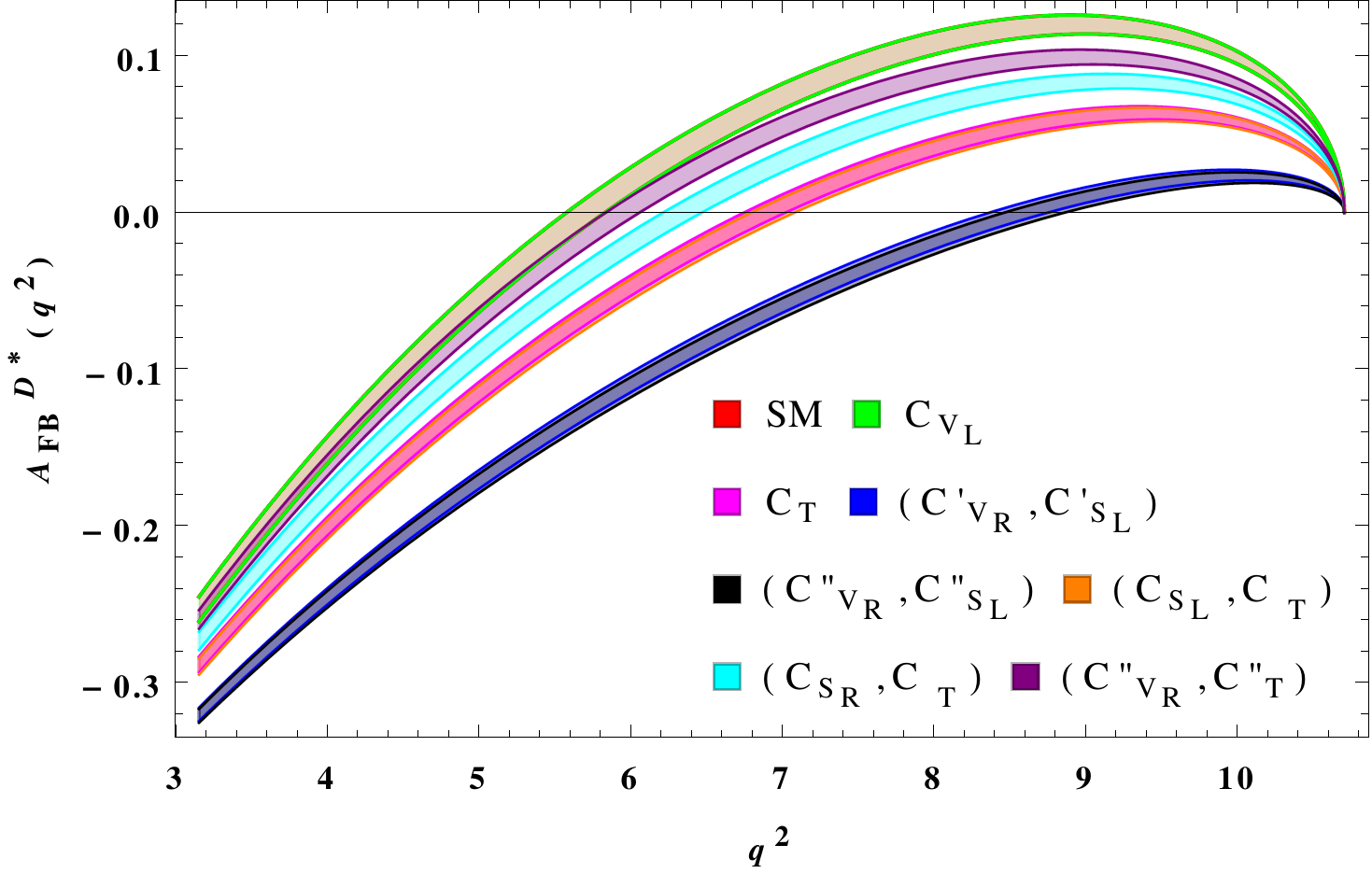}
\caption{Figure corresponds to $A_{FB}(q^2)$ as a function of $q^2$ for the $B\rightarrow D^*\tau\bar{\nu}$ decay. The band, representing 1$\sigma$ range, is mainly due to the uncertainties in various hadronic form factors and is obtained by adding these errors in quadrature.}
\label{fig1}
\end{figure}
%%%%%%%%%%%%%%%%%%%%%%%%%%%%%

From table~\ref{tab2} we find the following discriminating features among the allowed solutions:
\begin{itemize}
\item \underline{$\mathcal{O}_{V_L}$ and $\mathcal{O}_T$ solutions}: The $\mathcal{O}_{V_L}$ and $\mathcal{O}_T$ solutions predict $P_{\tau}^D \approx 0.35$ whereas all the other solutions predict it to be about $0.45$. Therefore a measurement of this observable to a precision of $0.1$ can distinguish these two solutions from the other five. A distinction between the $\mathcal{O}_{V_L}$ and $\mathcal{O}_T$ solutions can be obtained by measuring $R_D$ to a precision of $0.01$, which can be achieved at Belle II~\cite{Kou:2018nap}.

\item \underline{$(\mathcal{O}''_{V_R}, \mathcal{O}''_{T})$ and $(\mathcal{O}'_{V_R},\, \mathcal{O}'_{S_L})$ solutions}: The $(\mathcal{O}''_{V_R}, \mathcal{O}''_{T})$ and $(\mathcal{O}'_{V_R},\, \mathcal{O}'_{S_L})$ solutions predict $A_{FB}^D$ to be 
$\sim -0.24$ whereas other five solutions predict it to be $\sim 0.33$. Establishing this variable to be negative will distinguish these two solutions from the others. 
 A clear distinction between these two solutions can be made  through the measurement of zero crossing point (ZCP) of $A_{FB}^{D^*}(q^2)$. In fig.~\ref{fig1}, we have plotted $A_{FB}^{D^*}(q^2)$ as a function of $q^2$.
From this figure, we note that the ZCP for the $(\mathcal{O}''_{V_R}, \mathcal{O}''_{T})$ and $(\mathcal{O}'_{V_R}, \mathcal{O}'_{S_L})$ solutions are $\sim 6.0$ GeV$^2$ and $\sim 8.6$ GeV$^2$, respectively.
A further discrimination can be made through the branching ratio of $\mathcal{B}(B_c\rightarrow \tau\,\bar{\nu})$, predicted to be $11\%$ by the $(\mathcal{O}''_{V_R}, \mathcal{O}''_{T})$ solution and $6\%$ by the $(\mathcal{O}'_{V_R},\, \mathcal{O}'_{S_L})$ solution, provided it is measured to a precision of about $2\%$.

\item \underline{The other three solutions}: The three solutions, $(\mathcal{O}_{S_L}, \mathcal{O}_T)$, $(\mathcal{O}_{S_R}, \mathcal{O}_T)$ and $(\mathcal{O}''_{V_R},\, \mathcal{O}''_{S_L})$, all predict the same values for $P_{\tau}^D$ and $A_{FB}^D$.   A distinction between these three solution can be done by the measuring the ZCP of $A_{FB}^{D^*}(q^2)$ or by $\mathcal{B}(B_c\rightarrow \tau\,\bar{\nu})$. The ZCP of $A_{FB}^{D^*}(q^2)$ is $\sim 6.8$ GeV$^2$ for the $(\mathcal{O}_{S_L}, 
\mathcal{O}_T)$ and $(\mathcal{O}_{S_R}, \mathcal{O}_T)$ solutions and it is $\sim 8.6$ GeV$^2$ for the $(\mathcal{O}''_{V_R}, \mathcal{O}''_{S_L})$ solution. The respective predictions for $\mathcal{B}(B_c\rightarrow \tau\,\bar{\nu})$ of these three solutions are $0.8\%$, $4.0\%$ and $8.4\%$. Thus a measurement of  $\mathcal{B}(B_c\rightarrow \tau\,\bar{\nu})$ to a precision of $2\%$ can distinguish between these three solutions. 
\end{itemize}
 
\section{Conclusions}
After Moriond 2019, the discrepancy between the the global average values and the SM predictions of $R_D$-$R_{D^*}$ reduces to $3.1~\sigma$. The measured value of $f_L^{D^*}$ rules out the previously allowed tensor NP solution at $\sim 5\sigma$ level. We redo the fit with the new global averages and find that there are only {\bf seven} allowed NP solutions. We discuss methods to discriminate between these solutions by angular observables in $B \to (D,D^*) \,\tau\, \bar{\nu}$ decays and the branching
ratio $\mathcal{B}(B_c\rightarrow \tau\,\bar{\nu}$). We find that each of these seven solutions can be uniquely identified by the combination of the five observables with the following described precision:
(i) The $\tau$ polarization $P_{\tau}^D$ in $B\rightarrow D\, \tau \,\bar{\nu}$ to a precision $0.1$, (ii) The ratio $R_D$ to a precision of $0.01$, (iii) The $A_{FB}^D$ in $B\rightarrow D\, \tau\, \bar{\nu}$ to be either positive or negative, (iv) The zero crossing point of $A^{D^*}_{FB}(q^2)$ in $B\rightarrow D^*\,\tau\,\bar{\nu}$ to a precision of $0.5$ GeV$^2$, and (v) The branching ratio of $B_c\rightarrow \tau\,\bar{\nu}$ to a precision of $2\%$.
 
\emph{Acknowledgement}: SK thanks to the organizers for financial support to attend the workshop.

\input{referenc}

\end{document}

%% file: referenc.tex
%%%%%%%%%%%%%%%%%%%%%%%% referenc.tex %%%%%%%%%%%%%%%%%%%%%%%%%%%%%%
% sample references
% %
% Use this file as a template for your own input.
%
%%%%%%%%%%%%%%%%%%%%%%%% Springer-Verlag %%%%%%%%%%%%%%%%%%%%%%%%%%
%
% BibTeX users please use
% \bibliographystyle{}
% \bibliography{}
%